# Robust Design of $H_\infty$ Controller for a Launch Vehicle Autopilot against Disturbances

A. Graells, F. Carrabina

*Department of Electrical Engineering, School of Industrial Engineering, Technical University of Madrid, Madrid, Spain*

***Abstract:*** *Atmospheric flight phase of a launch vehicle is utilized to evaluate the performance of an $H_\infty$ controller in the presence of disturbances. Dynamics of the vehicle is linearly modeled using time-varying parameters. An operating point was found to design a robust command tracker using $H_\infty$ control theory that guarantees a stable maneuver. At the end, the controller was employed on the launch vehicle to assess the capability of control design on the linearized aerospace vehicle. Experimental results illustrate the excellent performance of the $H_\infty$ controller and accurate tracking implemented by the autopilot. Also the robustness of the entire system against disturbances is demonstrated to be acceptable.*

## I. Introduction

It is obvious that when an aerospace launch vehicle (ALV) is launched, it has a mission which must be accomplished. To reach this aim, ALV has to move along a specified trajectory in order to reach a required attitude. These tasks are executed by the flight control system called autopilot. It provides the ALV with control actions such as forces and torques in order to achieve best fulfillment of the designed requirements to the vehicle terminal state vector during its powered path [1].

Since ALV is a dynamical system and this feature causes inaccurate and time-varying parameters, it can only be modeled in mathematical terms. So the most important problem arises as a result of the variable characteristics of such vehicles [2]. Therefore the attitude control systems are facing dynamics with uncertain parameters in addition to nonlinearities and disturbances. In order to achieve an acceptable performance, many nonlinear or optimal control methods have been proposed to follow the nominal trajectory [3, 4].

When ALV is doing its mission in the real environment, many disturbances in different forms affect on ALV's performance and its accuracy. So a large number of control problems of practical importance which can be described using the block diagram shown in Figure. 1 involves designing a controller capable of stabilizing a given system while minimizing the worst-case response to exogenous disturbances. This problem is relevant, for instance, for disturbance rejection and tracking. It is known that $H_\infty$ control [5, 6] is an effective method for attenuating such disturbances [7, 8] and can be applied on vehicles requiring robust controllers such as flight stabilizer [9], VSTOL aircraft [10], and remote passive acoustic monitoring (PAM) vehicles that are used for data collection, call classification, and species identification [11, 12]. In this theory the exogenous disturbances are modeled as bounded signals and performance is measured in terms of the resulted output. On the other hand the design objective is to synthesize an internally stabilizing controller $K(s)$ such that the energy of the output is kept below a given level [13]. It should be noticed that finding an optimal controller is difficult and complicated both analytically and even practically because of numerical difficulties in computation of unusual properties. So this problem can be limited to getting a suboptimal controller whose results are acceptable and much closed to real one. More and detailed information can be found in [14].

In this research, at first an ALV dynamic system is introduced and then $H_\infty$ controller is designed for both linear time invariant (LTI) model and linear time varying (LTV) model. Finally, simulation is run in the presence of perturbations and results are analyzed.

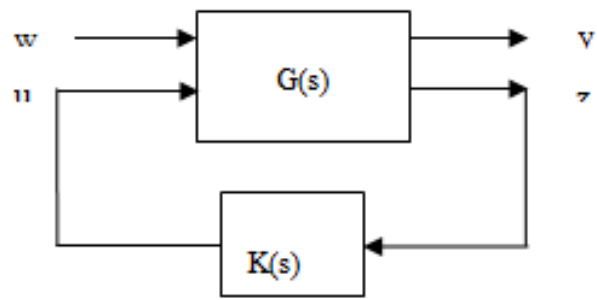

**Figure 1.** General model for $H_\infty$ problem

Section 2 presents $H_\infty$ control theory. ALV dynamic model is offered in Section 3. $H_\infty$ autopilot designed following simulation results demonstrated in section 4 and conclusion is presented in Section 5.

## II. H∞ Control Theory

In this section, linear $H_\infty$ control is briefly reviewed. Consider the following linear plant when the full state is available:

$$\dot{x} = Ax + Bu + B_w w \qquad (1)$$
$$y = Cx + Du$$

where $x \in R^n$ is the state vector, $u \in R^m$ is the control vector, $w \in R^p$ is the exogenous disturbance vector and, $y \in R^{n+m}$ is the output signal. Matrices C and D have suitable dimension and create a relationship between the cost of the state and the control's cost.

The $H_\infty$ control goal is to keep the closed-loop transfer matrix $T_{yw}$ (which goes from $w$ to $y$) below a given threshold $\gamma$. In other words

$$\|T_{yw}\|_\infty < \gamma \qquad (2)$$

*Assumption: (A,B) is stabilizable, (C,A) is detectable, and $D^T[C\ D]=[\ 0\ I\ ]$.*

The first assumption is necessary for system to be stabilized via output feedback. The second one guaranties the boundedness of the state. The last assumption means that $z$ has no cross weighting between the state and control and also the control weight matrix is the identity.

Since the goal is to attenuate disturbances, the $H_\infty$ norm of the transfer matrix $T_{yw}$ should be minimized by finding an appropriate controller. In other words, there exists a controller as:

$$u = -B^T X_\infty x \qquad (3)$$

where $X_\infty$ is a real, symmetric and positive semi-definite matrix which satisfies the Riccati equation:

$$XA + A^T X - X(BB^T - \gamma^{-2} B_w B_w^T)X + C^T C = 0 \qquad (4)$$

So the controller (3) is the optimal control in the sense that it minimizes the same quantity in the presence of the worse-case disturbance. Since for LTI stable systems induced norm in the time domain coincides with the $H_\infty$ norm of the transfer matrix in the frequency domain, the $H_\infty$ problem can be realized as disturbance attenuation [15] and therefore our goal is met.

## III. The Equations Of Motion For Alv

The governing equations of motion for an ALV can be derived from Newton's second law. Assuming rigid airframe for ALV, the 6DOF equations of motion obtained as follows [2]:

$$\begin{aligned}
F_x &= m(\dot{U} + qW - rV) \\
F_y &= m(\dot{V} + rU - pW) \\
F_z &= m(\dot{W} + pV - qU) \\
M_x &= I_x \dot{p} \\
M_y &= I_y \dot{q} + (I_x - I_y)pr \\
M_z &= I_z \dot{r} + (I_y - I_x)pq
\end{aligned} \qquad (5)$$

Since the ALV attitude control systems are usually designed based on the rendered linear equations of motion, a linear ALV is used in this article. Considering small perturbations, linearized equations of motion can be obtained as follows [1]:

$$\begin{aligned}
\dot{v}_z &= Z_v v_z + Z_q q + Z_\theta \theta + Z_{\delta e}\delta_e \\
\dot{q} &= M_{vz} v_z + M_q q + M_{\delta e}\delta_e \\
\dot{v}_y &= Z_v v_y + Z_r r + Z_\theta \theta + Z_{\delta r}\delta_r \\
\dot{r} &= M_{vy} v_y + M_r r + M_{\delta r}\delta_r \\
\dot{p} &= M_p p + M_{\delta a}\delta_a
\end{aligned} \qquad (6)$$

where *Z, M* are time-varying dynamic coefficients and *δ* is deflection of trust vector.

Since the control objective is to track guidance command in pitch channel, thus the two first equations will be regarded as required dynamics and the other three ones are waved belonging to yaw and roll channels. Time varying coefficients of pitch dynamics in Eq. (17) are shown as in Figure. 2.

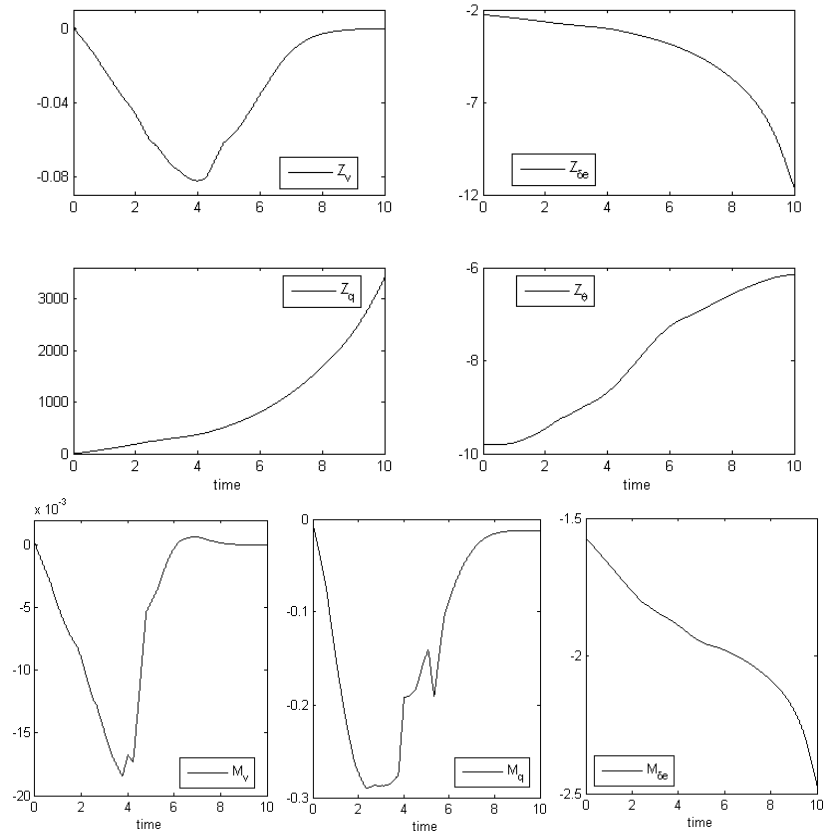

**Figure 2.** Longitudinal dynamic coefficient

The servo dynamics describing the trust vector deflection is:

$$[TF]_{servo} = \frac{\delta}{\delta_c} = \frac{1}{0.1s+1} \quad (7)$$

with a rate limit of $|\frac{d\delta}{dt}| < 25 \deg/\sec$.

Since the reference signal is pitch rate, a rate gyro is selected as follows:

$$[TF]_{gyro} = \frac{(80\pi)^2}{s^2 + (40\pi)s + (80\pi)^2} \quad (8)$$

### IV. Autopilot Design And Simulation Results

In this section, the $H_\infty$ control is designed for both LTI and LTV aerospace launch vehicle pitch longitudinal channel and related results are demonstrated.

**4.1 $H_\infty$ Control Design for LTV Plant**

Since the pitch rate trajectory is desired to be tracked, a new variable change is considered to meet our goal. So the pitch rate can be stated in terms of tracking error as:

$$q = q_c - e \quad (9)$$

where $q_c$ is the nominal pitch rate profile. So a new state vector is defined as:

$$\hat{x} = [\int e \quad e \quad v_z]^T$$

Now by substituting (9) into the longitudinal equations of (6), the time-varying state space equations of ALV transform into following form:

$$\hat{\dot{x}} = \begin{bmatrix} 0 & 1 & 0 \\ 0 & M_q & -M_v \\ -Z_\theta & -Z_q & Z_v \end{bmatrix} \hat{x} + \begin{bmatrix} 0 \\ -M_\delta \\ Z_\delta \end{bmatrix} u + \begin{bmatrix} 0 & 0 \\ 0 & 1 \\ 1 & 0 \end{bmatrix} w + \begin{bmatrix} 0 \\ \dot{q}_c - M_q q_c \\ Z_q q_c + Z_\theta \int q_c d\tau \end{bmatrix}$$

$$\hat{y} = \begin{bmatrix} 0 & 1 & 0 \end{bmatrix} \hat{x}$$

(11)

In this approach, in order to find the appropriate control, first the dynamic coefficients at the moment of 100 seconds are extracted as shown in Table 1.

**Table 1.** Dynamic coefficients at t = 100 sec

| | |
|---|---|
| $Z_v$ = -0.0020551 | $M_v$ = 0.0002725 |
| $Z_q$ = 1827.8 | $M_q$ = -0.014108 |
| $Z_\theta$ = -6.4939 | $M_\delta$ = -2.1086 |
| $Z_\delta$ = -6.2007 | |

Then using these coefficients, matrices *A, B* of the ALV state equations (11) are obtained. So by assuming $\gamma = 7.8$ the Riccati equation (4) can be solved. The solution is achieved as following:

$$X_\infty = \begin{bmatrix} 63.3031 & 0.6819 & 0.034 \\ 0.6819 & 1.8298 & -0.0002 \\ 0.034 & -0.0002 & 0.0000 \end{bmatrix}$$ (12)

Substituting into Eq. (3) the $H_\infty$ controller is found as:

$$u_{\text{var}} = -\begin{bmatrix} 0 \\ -M_\delta \\ Z_\delta \end{bmatrix} \begin{bmatrix} 63.3031 & 0.6819 & 0.034 \\ 0.6819 & 1.8298 & -0.0002 \\ 0.034 & -0.0002 & 0.0000 \end{bmatrix} \hat{x}$$ (13)

**4.2 $H_\infty$ Control Design for LTI Plant**

In order to compare the performance of the $H_\infty$ controller in both LTV and LTI systems, a linear time invariant ALV model is considered and the appropriate controller is designed.

As the first step, dynamic coefficients at the moment of 60 seconds from the time-varying system are extracted as depicted in Table 2.

**Table 2.** Dynamic Coefficients at t = 60 sec

| | |
|---|---|
| $Z_v$ = -0.054252 | $M_v$ = -0.003439 |
| $Z_q$ = 608.84 | $M_q$ = -0.18404 |
| $Z_\theta$ = -6.4939 | $M_\delta$ = -1.9594 |
| $Z_\delta$ = -3.4855 | |

So the LTI model with a structure like (11) is established which guarantees the stability of the whole system. Putting constant matrices of plant into the Eq. (4) leads to solving it. The matrix

$$X_\infty = \begin{bmatrix} 25.4427 & 0.7938 & 0.0405 \\ 0.7938 & 0.809 & 0.0013 \\ 0.0405 & 0.0013 & 0.0001 \end{bmatrix}$$ (14)

is as a solution for Eq. (4) when $\gamma = 20$. Thus the resulting controller is obtains as:

$$u_{in\,\text{var}} = -[1.4141 \quad 1.5804 \quad 0.0024]\,\hat{x}$$ (15)

*Robust Design of $H_\infty$ Controller for a Launch Vehicle Autopilot against Disturbances*

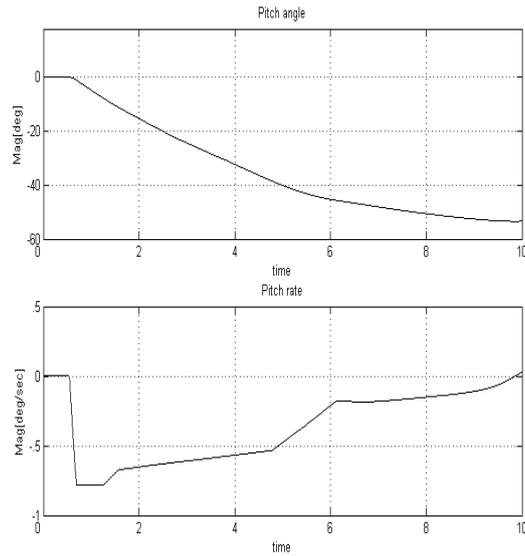

**Figure 3.** Desired pitch and pitch rate to be tracked

In this paper, pitch rate program has been designed offline as shown in Fig. 3 and it is desired to be tracked during the flight envelope. Simulation was run in presence of disturbances as unmatched and depicted in Fig. 4. The simulation results with $H_\infty$ control for LTV and LTI systems are illustrated in Fig. 5 and Fig. 6, respectively.

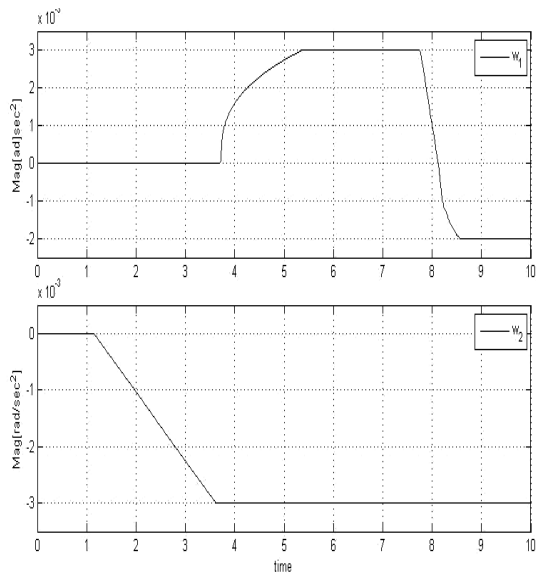

**Figure 4.** Exogenous disturbances profiles

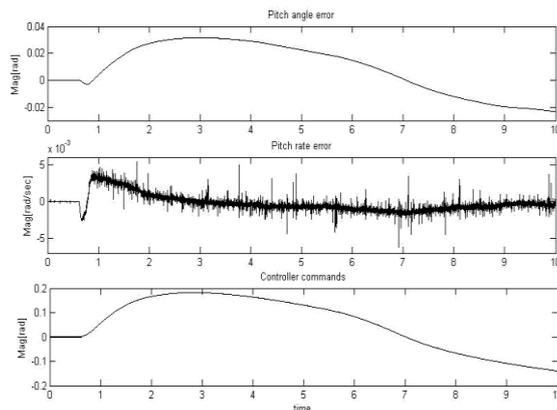

**Figure 5.** Pitch angle, pitch rate realization error and controller command for LTI plant

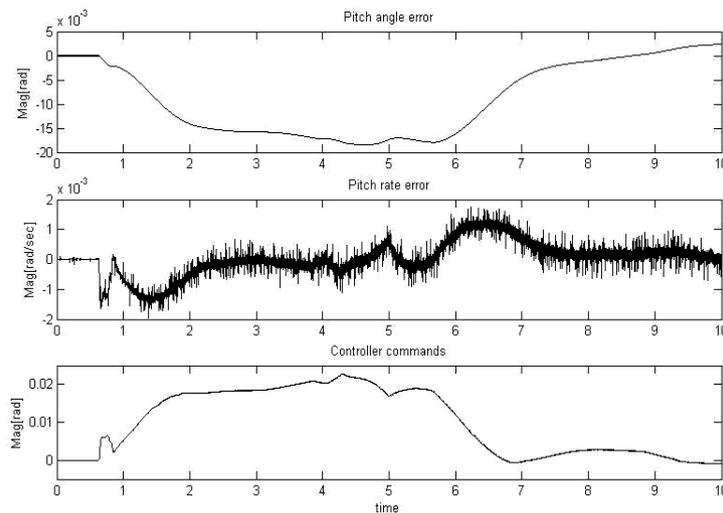

**Figure 6.** Pitch angle, pitch rate realization error and controller command for LTV plant

## V. Conclusion

In this research, commanded pitch rate tracking with exogenous disturbances for the atmospheric flight of a time-varying ALV is considered based on $H_\infty$ control approach. Also a LTI aerospace launch vehicle model is extracted for comparison purpose. Results show that $H_\infty$ controller performance in LTV system against the unmatched disturbances exhibits much desirable performance while pitch rate and pitch angle error in LTI system are noticeable and nominal trajectories are not properly tracked.

## References

[1] V. V. Malyshev, *Aerospace vehicle Control: Modern Theory and Applications* (Sao Paulo, Brazil: IAE, 1996).
[2] J. H. Blacklock, *Automatic Control of Aircraft and Missiles* (New York: Wiley, 1991).
[3] V. V. Malyshev, and M. N. Krasilshikov, *Aerospace vehicle Navigation and Control* (Sao Paulo, Brazil: FAPESP, 1996).
[4] M. Esfahanian, B. Ebrahimi, J. Roshanian, and M. Bahrami, Time-Varying Dynamic Sliding Mode Control of a Nonlinear Aerospace Launch Vehicle, *Canadian Aeronautics and Space Journal*, 57(2), 2011.
[5] A. Stoorvogel, *The $H_\infty$ Control Problem: A State Space Approach* (New York: Prentice-Hall, 1992).
[6] J. K. Doyle, P. P. Glover, Khargonekar and B. A. Frances, State Space Solution to Standard $H_\infty$ Control Problems, *IEEE Trans. Automat. Contr.* 34(8), 1989, 831-847.
[7] R. L. Pereira and K. H. Kienitz, Tight formation flight control based on $H_\infty$ approach, *Proc. 2016 24th Mediterranean Conference on Control and Automation (MED)*, Athens 2016, 268-274.
[8] D. Ye, and G.-H. Yang, *Delay-dependent adaptive reliable $H_\infty$ control of linear time-varying delay systems*. Int. J. Robust Nonlinear Control, Vol. 19, 2009, 462–479.
[9] Ş. Akyürek, G. S. Özden, B. Kürkçü, Ü Kaynak and C. Kasnakoğlu, Design of a flight stabilizer for fixed-wing aircrafts using $H_\infty$ loop shaping method, *Proc. 2015 9th International Conference on Electrical and Electronics Engineering (ELECO)*, Bursa, 2015, 790-795.
[10] R. A. Hyde and K. Glover, The application of scheduled $H_\infty$ controllers to a VSTOL aircraft, in *IEEE Transactions on Automatic Control*, 38(7), pp. 1021-1039, Jul 1993.
[11] M. Esfahanian, H. Zhuang, and N. Erdol, *Using Local Binary Patterns for Classification of Dolphin Calls*, Journal of Acoustical Society of America, 134(1), 2013.
[12] M. Esfahanian, H. Zhuang, and N. Erdol, *A New Approach for Classification of Dolphin Whistles*, 2015 IEEE International Conference on Acoustics, Speech, and Signal Processing (ICASSP), Florence, Italy, 2014, 6038-6042.
[13] R. S. Sanchez-Pena, and M. Sznaier, *Robust Systems: Theory and Applications* (New York: Wiley, 1998).
[14] Q. ZHANG, S. YE, Y. LI, and X WANG, An Enhanced LMI Approach for Mixed H2/H∞ Flight Tracking Control, *Chinese Journal of Aeronautics*, 24(3), 2011, 324-328.
[15] F. Castanos, and L. Fridman, Robust Design Criteria for Integral Sliding surfaces, *Proc. 44th IEEE Conference on Decision and Control*, Seville, Spain, 2005, 1976-1981.